\documentclass[12pt]{article}
\def\cdate{{February 13, 2025}}
\usepackage{amsfonts}
\usepackage{amsmath}
\usepackage{latexsym}
\usepackage{amssymb}
\usepackage{txfonts}
\usepackage{bm}
\usepackage[american]{babel}
\usepackage{graphicx}
\usepackage{framed}
\usepackage{xcolor}
\definecolor{mygray}{gray}{0.95} 
\definecolor{mydarkgray}{gray}{0.70} 
\colorlet{shadecolor}{mygray}
\makeatletter
\def\timenow{%
\@tempcnta=\time \divide\@tempcnta by 60 \number\@tempcnta:\multiply
\@tempcnta by 60 \@tempcntb=\time \advance\@tempcntb by -\@tempcnta
\ifnum\@tempcntb <10 0\number\@tempcntb\else\number\@tempcntb\fi}
\newcounter{outputpage}
\renewcommand{\@oddhead}
{\stepcounter{outputpage}\hfill\hfill\theoutputpage}
\renewcommand{\@evenhead}
{\stepcounter{outputpage}\hfill\hfill\theoutputpage}
\renewcommand{\@oddfoot}
{\vbox{
\hrule
\vspace{3pt}
\hfil
{\scriptsize\textit{
\hfill\hfill\jobname.tex; \today; \timenow; p. \theoutputpage}}
\hfil
}}
\renewcommand{\@evenfoot}
{\vbox{
\hrule
\vspace{3pt}
\hfil
{\scriptsize\textit{
\hfill\hfill\jobname.tex; \today; \timenow; p. \theoutputpage
}}
\hfil
}}
\makeatother

\def\nmt{{
\null
\vspace{-4cm}
\par
\hspace*{50truemm}{\hrulefill}
\par
\vskip-4truemm
\par
\hspace*{50truemm}{\hrulefill}
\par\vskip5mm
\par
\hspace*{50truemm}{{\large\sc
New Mexico Tech {\rm
(\cdate)
}}}\vskip4mm
\par
\hspace*{50truemm}{\hrulefill}
\par
\vskip-4truemm
\par
\hspace*{50truemm}{\hrulefill}
\par}}

\def\RR{\mathbb{R}}

\def\cD{\mathcal{D}}

\def\cH{\mathcal{H}}

\def\cS{\mathcal{S}}

\def\dim{\mathrm{dim\,}}

\def\Ind{\mathrm{Ind\,}}

\def\Ker{\mathrm{Ker\,}}

\def\supp{\mathrm{supp\,}}
\def\tr{\mathrm{tr\,}}
\def\Tr{\mathrm{Tr\,}}
\def\vol{\mathrm{vol\,}}

\def\la{\langle}
\def\ra{\rangle}
\def\proof{\par\noindent{\textbf{\textit{Proof:\;}}}}

\def\nn{{\nonumber}}
\def\be{\begin{equation}}
\def\ee{\end{equation}}
\def\bea{\begin{eqnarray}}
\def\eea{\end{eqnarray}}
\def\bed{\begin{definition}{\ }}
\def\eed{\end{definition}}
\def\ed{\end{document}}
\def\bp{\begin{proposition}}
\def\ep{\end{proposition}}
\def\bc{\begin{center}}
\def\ec{\end{center}}
\def\bi{\begin{itemize}}
\def\ei{\end{itemize}}
\def\benum{\begin{enumerate}}
\def\eenum{\end{enumerate}}
\def\bmp{\begin{minipage}}
\def\emp{\end{minipage}}

\newtheorem{lemma}{Lemma}

\newtheorem{proposition}{Proposition}

\newtheorem{definition}{Definition}
\begin{document}

\begin{titlepage}
\thispagestyle{empty}
\nmt

\bigskip
\bigskip
\bigskip
\bigskip
\centerline{\LARGE\bf On Zero Energy States}
\bigskip
\centerline{\LARGE\bf in SUSY Quantum Mechanics}
\bigskip
\centerline{\LARGE\bf on Manifolds}
\bigskip
\bigskip
\bigskip
\centerline{\Large\bf Ivan G. Avramidi}
\bigskip
\centerline{\it Department of Mathematics}
\centerline{\it New Mexico Institute of Mining and Technology}
\centerline{\it Socorro, NM 87801, USA}
\centerline{\it E-mail: ivan.avramidi@nmt.edu}
\bigskip
\medskip

\begin{abstract}
We study the zero modes of the operator $H_f{}{}=D^*_fD_f{}{}$,
with a Dirac type operator $D_f{}{}$, acting on the spinor bundle
over a closed even dimensional Riemannian manifold $M$.
The operator $D_f{}{}=D+ifI$ is a deformation of the Dirac operator
$D$ by a smooth function $f$.
We obtain sufficient conditions on the deformation function that
guarantee the positivity of the operator $H_f{}{}$, that is,
the absence of zero modes. We also show that these conditions
are not necessary and provide an explicit counterexample of a zero
mode of the operator $H_f{}{}$.

\end{abstract}

\end{titlepage}


\section{Introduction}
\setcounter{equation}{0}

This paper was initially motivated by the question, whether certain
supersymmetric matrix models possess normalizable zero-energy states
\cite{froehlich00,froehlich98, graf01}. These models are supersymmetric
extensions of bosonic membrane matrix models and were studied as reduced
supersymmetric Yang Mills theories and as super-membrane matrix models. The
question boils down to the study of the spectrum of a
supersymmetric Hamiltonian acting on
vector valued functions in $\RR^2$,
\be
\tilde H=-I\Delta + V,
\label{11iga}
\ee
where $I$ is the unit matrix, $\Delta=\partial_x^2+\partial_y^2$ is the Laplacian,
and $V$ is the matrix potential
\be
V=
\left(
\begin{matrix}x^2y^2+x &y\\{}
	y & x^2y^2-x\\
\end{matrix}\right).
\ee
This Hamiltonian is equal to
\be
\tilde H = \tilde D^* \tilde D,
\ee
where $\tilde D$ is the operator
\be
\tilde D=i\left(
\begin{matrix}
\partial_y -xy & \partial_x\\{}
\partial_x & -\partial_y - xy\\
\end{matrix}\right).
\ee
Similar problems were studied by Simon \cite{simon83}, and Fefferman and Phong
\cite{fefferman81}, also see \cite{cycon08}.

In this paper we study a more general problem by considering the Hamiltonian
$H_f{}=D^*_fD_f$ of a deformed Dirac operator $D_f{}{}=D+ifI$, with an arbitrary
smooth function $f$, on a closed Riemannian manifold $M$. We find some sufficient
conditions on the function $f$ such that the operator $H_f{}{}$ is strictly
positive. We also construct a counterexample, that is, a special manifold $M$
and a function $f$ such that the operator $H_f{}{}$ has a normalized zero mode.

In Sec. 2 we briefly describe the algebra of the supersymmetric quantum mechanics
and the Witten index.
In Sec. 3 we describe the construction of the Dirac operator $D$
on Riemannian manifolds in the form suited for our study.
In Sec. 4 we introduce a deformation $D_f{}{}$ of the Dirac operator by a smooth function
$f$. In Sec. 5 we consider a two-dimensional example and show that
for a specific function $f$ it leads to the Hamiltonian (\ref{11iga})
on the Euclidean plane $\RR^2$.
In Sec. 6 we prove some sufficient conditions for the absence of zero modes
of the deformed Dirac operator $D_f{}{}$ (and, therefore, for the
positivity of the corresponding Hamiltonian $H_f{}{}$).
In Sec. 7 we prove various properties of the zero modes
and in Sec. 8 we provide a specific example of such a zero mode on a product
manifold $M=N\times S^1$. In Sec. 9 we briefly summarize our results.

\section{Supersymmetric Quantum Mechanics}
\setcounter{equation}{0}

We review briefly the supersymmetric quantum mechanics in the form adopted
to our needs (for more details, see \cite{tong}).
The supersymmetric quantum mechanics is described by a
self-adjoint involution
$J$ and a nilpotent operator $Q$, called the supercharge,
on a Hilbert space $\cH$ satisfying the algebra
\be
J{}{}^2=I,
\qquad
J^*=J,
\qquad
Q^2=0,
\qquad
J{} Q=-QJ{}{}=-Q.
\label{22iga}
\ee
The  supersymmetric Hamiltonian is defined by
\be
H=(Q+Q^*)^2=H_++H_-,
\ee
where
\be
H_+=Q^*Q,
\qquad
H_-=QQ^*,
\ee
First, by using the orthogonality of the operators $H_+$ and $H_-$,
$H_+H_-=H_-H_+=0$,
we get
\bea
\Tr\exp(-tH)
&=&\Tr\left\{ \exp(-tH_+)+\exp(-tH_-)\right\}.
\\
\Tr J{}{}\exp(-tH)
&=&\Tr\left\{\exp(-tH_+)-\exp(-tH_-)\right\}.
\eea
Next, by using the intertwining relations
\be
QH_+=H_-Q,
\qquad
H_+Q^*=Q^*H_-,
\label{213iga}
\ee
we obtain
\bea
Q\exp(-tH_+)Q^* &=&H_-\exp(-tH_-),
\\
Q^*\exp(-tH_-)Q &=& H_+\exp(-tH_+).
\eea
and, therefore,
\be
\Tr H_+\exp(-tH_+) =\Tr H_-\exp(-tH_-).
\ee
This leads to a nontrivial property
\be
\frac{d}{dt} \Tr J{}{}\exp(-tH)
=-\Tr \left\{H_+\exp(-tH_+) - H_-\exp(-tH_-)\right\}=0,
\ee
which means that the quantity
\be
\Ind Q=\Tr J{}{}\exp(-tH)
\ee
does not depend on $t$.
It defines the
index of the operator $Q$,
\be
\Ind Q
=\dim\Ker H_+-\dim\Ker H_-,
\ee
which, in the context of supersymmetry, is called the Witten index.

In a special situation when the adjoint supercharge operator $Q^*$
satisfies the intertwining relation
\be
\Gamma Q=\pm Q^*\Gamma,
\ee
with a self-adjoint involution $\Gamma$,
the Hamiltonians also satisfy such a relation
\be
\Gamma H_+ = H_-\Gamma.
\ee
This means that the operators $H_+$ and $H_-$ have the
same spectrum (including the kernels), in particular,
\be
\Tr\exp(-tH_+)=\Tr\exp(-tH_-),
\ee
and, therefore, the index vanishes, $\Ind Q=0$.

Obviously, the operators $H_+$, $H_-$ and $H$ are non-negative by construction.
They  are {\it almost}
isospectral, that is, they have the same positive spectrum
and the only difference is in the number of zero modes.
The supersymmetry is said to be broken if the Hamiltonian $H$
is strictly positive and unbroken if it has a zero mode.
Therefore, if the index is non-zero, then there must be some zero
modes and the supersymmetry is not broken. However, if the index is
equal to zero, it does not mean that there are no zero modes.
It just means that the number of zero modes of the operators $H_+$ and
$H_-$ are equal.


\section{Dirac Operator}
\setcounter{equation}{0}

In this section we follow our paper \cite{avramidi05} (for more details see this
paper). Let $(M,g)$ be a smooth compact Riemannian spin manifold of even
dimension $n=2m$ without boundary, equipped with a positive definite Riemannian
metric $g$. 
We denote the local coordinates on $M$ by $x^\mu$, with Greek indices running
over $1,\dots, n$. The Riemannian volume element is defined as usual by
$d\vol=dx\,g^{1/2}\,,$ where $g=\det g_{\mu\nu}\,$ and and $dx=dx^1\wedge
\dots\wedge dx^n$ is the standard Lebesgue measure.

Let ${\cal S}$ be the spinor bundle over the manifold $M$ equipped with a
Hermitian fiber inner product $\la \;,\;\ra$. This naturally identifies the dual
vector bundle ${\cal S}^*$ with ${\cal S}$. The fiber inner product on the
spinor bundle ${\cal S}$ and the fiber trace, $\tr$, defines the natural $L^2$
inner product $(\;,\;)$ and the $L^2$-trace, $\Tr$, using the invariant
Riemannian measure on the manifold $M$. The completion of the space
$C^\infty(\cS)$ of smooth sections of the spinor bundle ${\cal S}$ in this norm
defines the Hilbert space $L^2({\cal S})$ of square integrable sections.


Let $\partial_\mu$ be the coordinate basis for the
tangent space $T_xM$ at a point $x\in M$. We use Latin indices from the
beginning of the alphabet, $a,b,c,d,\dots,$ to denote the frame components, they
also range over $1,\dots, n$. Let $ e_a=e_a{}^\mu\partial_\mu $ be an orthonormal
basis for the tangent space $T_xM$ so that
\be
g^{\mu\nu}=\delta^{ab}e_a{}^\mu e_b{}^\nu,
\qquad
g_{\mu\nu}e_a{}^\mu e_b{}^\nu=\delta_{ab}\,,
\ee
and $\sigma^a{}_\mu$
be the inverse transpose matrix to $e_a{}^\mu$, defining the dual basis $
\sigma^a=\sigma^a{}_\mu dx^\mu\, $ in the cotangent space $T_x^*M$, so that
\be
g_{\mu\nu}=\delta_{ab}\sigma^a{}_\mu\sigma^b{}_\nu,
\qquad
\qquad
g^{\mu\nu}\sigma^a{}_\mu \sigma^b{}_\nu=\delta^{ab}\,,
\ee
The spin connection one-form is defined by
\be
\omega_{abc}=
\frac{1}{2}
\left\{d\sigma_b(e_a,e_c)
-d\sigma_a(e_b,e_c)
+d\sigma_c(e_a,e_b)
\right\}.
\ee

We describe briefly the algebra of the Dirac matrices, for details see
\cite{zhelnorovich19,avramidi05}. The Dirac matrices $\gamma_{a}$,
$a=1,\dots,n$, are complex $2^m\times 2^m$ matrices forming a representation of
the Clifford algebra
\be
\gamma_{a}\gamma_{b}+\gamma_b\gamma_a=2\delta_{ab}I.
\ee
and the chirality operator $\Gamma{}$ is defined by
\be
\Gamma{}=\frac{i^{m}}{(2m)!}\varepsilon^{a_1\dots a_{2m}}\gamma_{a_1}\cdots\gamma_{a_{2m}}
=i^{m}\gamma_{1}\cdots\gamma_{2m}.
\ee
where $\varepsilon^{a_1\dots a_n}$ is the anti-symmetric Levi-Civita symbol.
We will use the basis in which all Dirac matrices are
Hermitian,
$
\gamma_{a}^{*{}}=\gamma_{a};
$
then the chirality operator is also Hermitian, $\Gamma^*=\Gamma$,
involutive
\be
\Gamma{}^2=I,
\ee
and anti-commutes with the Dirac matrices
\be
\Gamma{}\gamma_a=-\gamma_a \Gamma{}.
\ee
It defines the orthogonal projections
\be
P_\pm=\frac{1}{2}\left(I\pm \Gamma\right).
\ee
decomposing the spinor bundle into the left and right
spinors, $\cS=\cS_+\oplus \cS_-$.


The connection on the spinor bundle
$\nabla^{\cal S}: C^\infty({\cal S})\to
C^\infty(T^*M\otimes {\cal S}) $ defines the covariant
derivative in local coordinates
\be
\nabla_\mu\varphi
=\left(I\partial_\mu+{1\over 4}\gamma^{ab}\omega_{abc}\sigma^c{}_\mu\right)\varphi\,,
\ee
where
$
\gamma_{ab}=\gamma_{[a}\gamma_{b]}.
$
The connection is given its unique natural extension to bundles in the tensor
algebra over ${\cal S}$ and ${\cal S}^*$, and, using the Levi-Civita connection
of the metric $g$, to all bundles in the tensor algebra over ${\cal S},\,{\cal
S}^*,\,TM$ and $T^*M$. The commutator of the covariant derivatives is
\be
[\nabla_\mu,\nabla_\nu]\varphi
=\frac{1}{4}R_{\alpha\beta\mu\nu}\gamma^{\alpha\beta}\varphi,
\ee
where $R_{\alpha\beta\mu\nu}$ is the Riemann tensor and
$\gamma^{\mu\nu}=\gamma^{[\mu}\gamma^{\nu]}$ with
\be
\gamma^\mu=\gamma^ae_a{}^\mu.
\ee


The Dirac operator is a first order partial differential operator acting on
smooth sections of the spinor bundle $ D: C^\infty({\cal S})\to C^\infty({\cal
S}) $ defined by
\bea
D &=& i\gamma^c e_c{}^\mu \nabla_\mu
\,.
\eea

The Dirac operator $D$ is a self-adjoint elliptic operator acting on smooth
sections of spinor bundle over a compact manifold without boundary. It is well
known that the operator $D$ has a discrete real spectrum. Each eigenspace is
finite-dimensional and the eigenspinors are smooth sections of the spinor bundle
that form an orthonormal basis in $L^2({\cal S})$; for details, see
\cite{berline92,friedrich00,avramidi05}. One can show that for all non-zero
eigenvalues there is an isomorphism between the right and left eigenspaces. In
particular, their dimensions, that is, the multiplicities of the right and the
left eigenspinors corresponding to the same non-zero eigenvalue are equal. This
does not work for the zero eigenvalues; so there could be any number of right or
left eigenspinors corresponding to zero eigenvalue.

It is easy to show it has the form (which is known as the Lichnerowicz formula
\cite{friedrich00})
\be
D^2=-\Delta+\frac{1}{4}I R,
\ee
where
\be
\Delta=g^{\mu\nu}\nabla_\mu\nabla_\nu
\ee
is the spinor Laplacian and $R$ is the scalar curvature.
The square of the Dirac operator is obviously a non-negative operator
\be
(\varphi,D^2\varphi)
=||D\varphi||^2
=||\nabla\varphi||^2
+\frac{1}{4}(\varphi,R\varphi)\ge 0.
\ee
Therefore, if the second term is {positive} (that is, for positive scalar
curvature manifolds), then the square of the Dirac operator $D^2$ is strictly
positive, i.e. it does not have any zero modes.

The chirality operator anti-commutes with the Dirac operator,
\be
\Gamma D=-D\Gamma.
\ee
It plays the role of the involution $J=\Gamma$
in the supersymmetric quantum mechanics together
with the supercharge
\be
Q=P_-D=DP_+,
\qquad
Q^*=P_+D=DP_-.
\ee
The Hamiltonian is defined by
\be
H=D^2=H_++H_-,
\ee
where
\be
H_+=Q^*Q=P_+D^2,
\qquad
H_-=QQ^*=P_-D^2.
\ee

Therefore, one has, in particular,
\be
\Gamma D^2\exp(-tD^2)=-D\Gamma \exp(-tD^2)D,
\ee
and, hence,
\be
\frac{d}{dt}\Tr\Gamma\exp(-tD^2)
=-\Tr\Gamma D^2\exp(-tD^2)=\Tr D^2\Gamma \exp(-tD^2)=0,
\ee
which means that the modified heat trace does not depend on $t$
and is equal to the index of the Dirac operator
\be
\Ind D=\Tr \Gamma\exp(-tD^2)
=\Tr \left\{\exp(-tH_+)-\exp(-tH_-)\right\}.
\ee

The asymptotic expansion of the heat kernel diagonal
of the square of the Dirac operator
the well known form \cite{gilkey95,berline92,avramidi00}
\be
U_{D^2}(t;x,x)\sim (4\pi t)^{-n/2}\sum_{k=0}^\infty \frac{(-1)^k}{k!}t^k a_k(D^2;x);
\ee
therefore, the heat trace has the asymptotic as $t\to 0^+$
\be
\Tr\exp(-tD^2)\sim (4\pi t)^{-n/2}\sum_{k=0}^\infty \frac{(-1)^k}{k!}t^k A_k(D^2).
\ee
where
\be
A_k(D^2)=\int_M d\vol \tr a_{k}(D^2).
\ee
It is easy to see that  for $k\ne m=n/2$,
\be
\int_M d\vol \tr \Gamma a_{k}(D^2)=0;
\ee
therefore, in even dimensions,
\be
\Ind D=(4\pi)^{-m}\frac{(-1)^m}{m!}\int_M d\vol \tr \Gamma a_{m}(D^2),
\ee
and in odd dimensions the index vanishes, $\Ind D=0$.

However, it does not mean that the Dirac operator does not have any zero modes
in odd dimensions. It is easy to construct an odd-dimensional closed manifold
with zero modes. Let $\Sigma$ be an even-dimensional closed manifold with a
non-zero index of the Dirac operator, $\Ind D_\Sigma\ne 0$. Then the
odd-dimensional manifold $N=\Sigma\times S^1$ will have zero modes of the Dirac
operator, $\dim\Ker D_{N}>0$, even though the index is zero, $\Ind D_N=0$.


\section{Deformed Dirac Operator $D_f{}{}$}
\setcounter{equation}{0}

Let $f\in C^\infty(M)$ be a smooth real valued function on the manifold $M$.
We decompose it via
\be
f=\mu+\tau h,
\ee
where
\be
\mu=\frac{1}{\vol(M)}\int_M d\vol\; f
\ee
is the average value of the function $f$,
$\tau$ is a positive real parameter,
 and $h$ is a function that satisfies
\be
\int_M d\vol\; h=0;
\ee
and normalized by
then
\be
||h||^2=1.
\ee
Such function can always be represented, for example,
 by $h=\Delta \phi$, where $\phi$ is uniquely
determined by the function $h$.

We define the deformed Dirac operator $D_f{}{}: C^\infty(\cS)\to C^\infty(\cS)$ by
\be
D_f{}{}=D+iI f
\ee
with the adjoint
\be
D^*_f{}{}=D_{-f}{}=D-iI f.
\ee
To make a connection with the supersymmetric quantum mechanics
we introduce the involution and the supercharge operator
$J, Q: C^\infty(\cS)\oplus  C^\infty(\cS)\to  C^\infty(\cS)\oplus  C^\infty(\cS)$
acting the pairs of spinors
by
\be
J=\left(
\begin{matrix}
I & 0\\
& -I\\
\end{matrix}\right)
\ee
and
\be
Q=\left(
\begin{matrix}
0 & 0\\
D_f{} & 0\\
\end{matrix}\right),
\qquad
Q^*=\left(
\begin{matrix}
0 & D_f{}^*\\
0 & 0\\
\end{matrix}\right).
\ee

The operator $D_f{}$
satisfies the (anti)-commutation relations
\bea
\Gamma D_f{}{}+D_f{}{} \Gamma &=& 2 i\Gamma f{},
\label{38iga}\\
\Gamma D_f{}{}-D_f{}{} \Gamma &=& 2\Gamma D,
\label{39iga}
\eea
and the intertwining relation
\be
\Gamma D_f{}{}=-{}D^*_{f}{}\Gamma{}.
\label{425iga}
\ee

The supersymmetric Hamiltonian is now defined by
\be
H=H_++H_-,
\ee
where
\bea
H_+ = Q^*Q=\left(
\begin{matrix}
H_f & 0\\
0 & 0\\
\end{matrix}\right),
\qquad
H_- = QQ^*=\left(
\begin{matrix}
0 & 0\\
0 & H_{-f}\\
\end{matrix}\right),
\eea
where
\be
H_f{}{} = D_f{}^*D_f{}
= D^2+m_f{},
\ee
with
\be
m_f{} =i[D,f]+If^2{}=-\gamma^\mu \nabla_\mu f+If^2{}.
\ee
Note that when $f\ne 0$ this matrix has the form
\be
m_f{}=f^2\left(\gamma^\mu\nabla_\mu {1\over f}+I\right).
\ee

The Hamiltonian
 satisfies the intertwining relation.
\be
\Gamma H_f{}{}={}H_{-f}\Gamma{}.
\label{426iga}
\ee
By using the intertwining relations (\ref{425iga}) and (\ref{426iga})
we have
\be
\Gamma\exp(-tH_f{}{})=\exp(-tH_{-f})\Gamma,
\ee
and, therefore,
\be
\exp(-tH_{-f})=\Gamma\exp(-tH_f{}{})\Gamma,
\ee
which gives
\be
\Tr \exp(-tH_f{}{}) = \Tr \exp(-tH_{-f}{}{}),
\ee
Then the Witten index is
\be
\Ind Q =
\Tr\left\{\exp(-tH_f)
-\exp(-tH_{-f})\right\}
=0
\ee

It is easy to see that
the Hamiltonian $H_f$
commutes with the operator $\Gamma D_f$,
\be
\Gamma D_f{}\exp(-t H_f) = \exp(-tH_f) \Gamma D_f{};
\ee
this gives
\be
D_f{}\exp(-t H_f) = -\Gamma \exp(-tH_f) D^*_f{}\Gamma.
\ee
and, therefore,
\be
\Tr D_f{} \exp(-tH_{f})
=-\Tr D^*_{f}{}\exp(-tH_{f})
\ee
Now,
by using  (\ref{38iga})
and (\ref{39iga}) we obtain
\bea
\Tr D\exp(-tH_{f})
=0.
\eea

Our primary interest is the study of the
spectrum of the Hamiltonian $H_f{}{}$.
Let $\varphi_\lambda$ be an eigenspinor
of the Hamiltonian $H_f{}$
with an eigenvalue $\lambda^2$,
\be
H_f{}\varphi_\lambda=\lambda^2\varphi_\lambda.
\ee
Then the spinor
\be
\psi_{\lambda}=\Gamma\varphi_\lambda
\ee
is
an eigenspinor of the operator $H_{-f}$ with
the same eigenvalue $\lambda^2$,
\be
H_{-f}\psi_{\lambda}=\lambda^2\psi_{\lambda}.
\ee
Therefore, the spectrum of the operator $H_f{}{}$ does not
depend on the {sign} of the function $f$.
So, there is an isomorphism between the eigenspaces of the operators $H_f{}{}$ and
$H_{-f}$ given just by the chirality operator, that is, for any $\lambda$,
\be
\Ker(H_f-\lambda^2 I)=
\Ker(H_{-f}-\lambda^2 I).
\ee

We define the functional
\bea
S_f{}(\varphi)&=&(\varphi,H_f{}{}\varphi)
=||Q_f\varphi||^2
=
||D\varphi||^2+M_f{}(\varphi)
\label{319iga}
\eea
where
\be
M_f{}(\varphi)=(\varphi,m_f{}\varphi).
\ee
In more details, it has the form
\be
S_f(\varphi)
=\int_M d\vol
\left\{|\nabla\varphi|^2
+\frac{1}{4}R|\varphi|^2
+\langle\varphi,m_f{}\varphi\rangle
\right\}.
\ee
Since this functional is non-negative
$
S_f{}(\varphi)\ge 0,
$
the spectrum of the operator $H_f{}{}$ is
{non-negative}.


We define the heat traces
\bea
\Theta(t,\mu,\tau) &=& \Tr\exp(-tH_f{}{}),
\\
\Psi(t,\mu,\tau) &=& \Tr\Gamma\exp(-tH_f{}{}),\textbf{}
\eea
Since the operator $H_f{}{}$ is non-negative the asymptotics as $t\to\infty$ of
the heat traces $\Theta(t,\mu,\tau)$ and $\Psi(t,\mu,\tau)$ depend on
the presence of the zero modes. Let
$P_0$ be the projection operator to the kernel $\Ker H_f{}{}$ (which is a
finite-dimensional vector space).
If there is a non-trivial kernel
of the Hamiltonian,
then as $t\to \infty$, the heat traces approach constants,
\bea
\Theta(t,\mu,\tau) &\sim &  \Tr P_0(\mu,\tau) +\cdots,
\\
\Psi(t,\mu,\tau) &\sim &  \Tr \Gamma P_0(\mu,\tau) +\cdots,
\eea
and, if the Hamiltonian $H_f{}{}$ is positive then these
traces are exponentially small, $\sim \exp(-t\lambda^2_1)$,
with $\lambda_1^2$ the bottom eigenvalue.

The asymptotic expansion of the heat traces as $t\to 0^+$ is determined by
the asymptotic expansion of the heat kernel diagonal
\cite{gilkey95,avramidi00,berline92}
\be
U_{H_f{}{}}(t;x,x)\sim (4\pi t)^{-n/2}\sum_{k=0}^\infty
\frac{(-1)^k}{k!}t^k a_k(H_f{}{};x),
\ee
and has the form
\bea
\Theta(t,\mu,\tau) &\sim& (4\pi)^{-m}\sum_{k=0}^\infty \frac{(-1)^k}{k!}t^{k-m}
A_k(\mu,\tau{}{}).
\\
\Psi(t,\mu,\tau) &\sim& (4\pi)^{-m}\sum_{k=0}^\infty \frac{(-1)^k}{k!}t^{k-m}
B_k(\mu,\tau{}{}).
\eea
where
\bea
A_k(\mu,\tau{}{}) &=&\int_M d\vol \tr a_{k}(H_f{}{}).
\\
B_k(\mu,\tau{}{}) &=&\int_M d\vol \tr \Gamma a_{k}(H_f{}{}).
\eea
The heat kernel coefficients $a_k(H_f)$
are  differential polynomials
in the function $f$, therefore, they are polynomials in
the parameters $\mu$ and $\tau$
satisfy the intertwining
relation
\be
a_k(H_{-f})=\Gamma a_k(H_f)\Gamma.
\ee
Therefore, the global coefficients are polynomials
satisfying
\be
A_k(-\mu,-\tau)=A_k(\mu,\tau),
\qquad
B_k(-\mu,-\tau)=B_k(\mu,\tau).
\ee

Notice that for $\mu=\tau=0$, the coefficients $A_k(0,0)$
are just the global heat kernel coefficients of the Dirac
operator,
\be
A_k(0,0)=A_k(D^2),
\ee
 all coefficients
$B_k(0,0)$ with $k\ne m$
vanish,
\be
B_k(0,0)=0,
\ee
and for $k=m$ it is equal to the index of the Dirac operator,
\be
B_m(0,0)=(-1)^m(4\pi)^m m!\Ind D.
\ee

It is easy to see that
in an  important case of a constant function $f=\mu$, that is, for $\tau=0$,
the Hamiltonian has the form
\be
H_f{}{}=D^2+\mu^2{}I,
\ee
and, therefore,
\bea
\Theta(t,\mu,0) & = & \exp(-t\mu^2{})\Tr\exp(-tD^2),
\\
\Psi(t,\mu,0) & = & \exp(-t\mu^2{})\Ind D.
\eea
Therefore, the heat kernel coefficients are
\be
A_k(\mu,0)=\sum_{j=0}^{k+m}{k\choose j}\mu^{2j} A_{k+m-j}(0,0),
\ee
The coefficients $B_k(\mu,0)$ vanish for $k=0,\dots, m-1$,
\be
B_k(\mu,0)=0,
\ee
and for $k\ge m$ are proportional to the index of the Dirac operator,
\be
B_k(\mu,0)=(-1)^{m}\frac{k!}{(k-m)!}(4\pi)^{m}\mu^{2(k-m)}\Ind D.
\ee


\section{Two-dimensional Manifolds}
\setcounter{equation}{0}

Let us restrict the above setup for the case of two-dimensional
manifolds, $n=2$.
The Dirac matrices are
\be
\gamma_1=\left(
\begin{matrix}0&-i\\{}
	i&0\\
\end{matrix}{}\right),\qquad
\gamma_{2}=\left(
\begin{matrix}0&1\\{}
	1&0
\end{matrix}\right),
\qquad
\Gamma{}=i\gamma_1\gamma_2=
\left(
\begin{matrix}1&0\\{}
	0&-1\\
\end{matrix}\right).
\ee
We denote
\be
\nabla_{(a)}=e_{(a)}^\mu\nabla_\mu,
\qquad
f_{(a)}=e_{(a)}^\mu\nabla_\mu f.
\ee
Then the Dirac operator $D$ has the form
\be
D = \left(
\begin{matrix}0& \nabla_{(1)}+i \nabla_{(2)}\\{}
	-  \nabla_{(1)}+i \nabla_{(2)} & 0\\
\end{matrix}\right),
\ee
Then the  deformed Dirac operator is
\be
D_f{}{} = \left(\
\begin{matrix} if{} & \nabla_{(1)} + i\nabla_{(2)}\\{}
	- \nabla_{(1)} +i \nabla_{(2)} & if{} \\
\end{matrix}\right),
\ee
and the Hamiltonian has the form
\be
H_f{}{} = -\Delta+I\left(\frac{1}{4}  R
+f^2{}\right)
+
\left(\begin{matrix} 0 & if_{(1)}-f_{(2)}\\{}
	-if_{(1)}-f_{(2)} & 0\\
\end{matrix}\right).
\ee

It is easy to show that by a unitary transformation of the Dirac matrices these
operators can be rewritten in the real (Majorana) form. By choosing
\be
T=\left(\begin{matrix}1&1\\{}
	i&-i\\\end{matrix}\right), \qquad
T^{-1}={1\over 2}\left(\begin{matrix}1&-i\\{}
	1&i\\\end{matrix}\right),
\ee
we have
\bea
\tilde\gamma_1 &=& T\gamma_1T^{-1}=\gamma_2,\qquad
\\
\tilde\gamma_2 &=& T\gamma_2T^{-1}=\Gamma,\qquad
\\
\tilde \Gamma{} &=& T\Gamma{}T^{-1} =\gamma_1.
\eea
By using these matrices we obtain the unitary equivalent operators
\bea
\tilde D &=& TDT^{-1}=
i\left(
\begin{matrix} \nabla_{(2)}& \nabla_{(1)}\\{}
	 \nabla_{(1)}&- \nabla_{(2)}\\
\end{matrix}\right),
\\[10pt]
\tilde D_f{}{} &=& TD_f{}{}T^{-1}
=i\left(
\begin{matrix}\nabla_{(2)} + f &  \nabla_{(1)} \\{}
	 \nabla_{(1)} &  -\nabla_{(2)} + f\\
\end{matrix}\right),
\\[10pt]
\tilde H_f{}{} &=& TH_f{}{}T^{-1}=-\Delta+I\left(\frac{1}{4} R
+f^2{}\right)
+\left(
\begin{matrix}-f_{(2)} &-f_{(1)}\\{}
	-f_{(1)} & f_{(2)}\\
\end{matrix}\right).
\eea

Of special interest is the torus $M=T^2=S^1\times S^1$
with the flat metric
$g_{\mu\nu}=\delta_{\mu\nu}$ and the function $f$ of the form
\be
f(x,y)=-\tau a^2 \sin\left(\frac{x}{a}\right) \sin\left(\frac{y}{a}\right),
\ee
where $a$ is the radius of the circles,
which, in the limit of infinite radius, $a\to \infty$, formally becomes the
Euclidean plane $M=\RR^2$ with the function
\be
f(x,y)=-\tau xy.
\ee
The Hamiltonian $H_f{}{}$ for this function takes the form
\be
H_f{}{} = I\left(-\Delta
+\tau^2x^2y^2\right)
+\tau \left(\begin{matrix} 0 &-iy+x\\{}
	iy+x & 0\\
\end{matrix}\right),
\ee
where
\be
\Delta=\partial_x^2+\partial_y^2,
\ee
and the operator $\tilde H_f{}{}$ is exactly the Hamiltonian
 (\ref{11iga}),
\be
\tilde H_f{}{} = I\left(-\Delta
+\tau^2 x^2y^2\right)
+\tau \left(
\begin{matrix}x &y\\{}
	y & -x\\
\end{matrix}\right).
\ee
These operators have been extensively studied in the literature in connection
with the reduced Yang-Mills theory, supersymmetric quantum mechanics, integrable
systems and others \cite{froehlich98,froehlich00,graf01}.


\section{Sufficient Condition for Positivity}
\setcounter{equation}{0}

We study the absolute
minimum of the functional $S_f{}(\varphi)$ (\ref{319iga}). In
particular, we study the minimizers $\varphi_*$ of this functional such that it
is equal to zero
\be
S_f{}(\varphi_*)=0;
\ee
obviously, $\varphi_*$ is a zero mode of the Hamiltonian $H_f{}{}$ (and of the
deformed Dirac operator $D_f{}{}$),
\be
H_f{}{}\varphi_*=D_f{}{}\varphi_*=0.
\ee

We have prove some sufficient conditions for the positivity of the
Hamiltonian.

\begin{proposition}
If one of the following conditions is valid
for any $\varphi$
\begin{enumerate}
	\item
	$M_f{}(\varphi)>0$,
	\item
	$M_f{}(\varphi)\ge 0$ and $D\varphi\ne 0$,
	\item
	$M_f{}(\varphi)>-\frac{1}{4}(\varphi,R\varphi)$,
	\item
	$M_f{}(\varphi)\ge -\frac{1}{4}(\varphi,R\varphi)$ and $\nabla\varphi\ne 0$,
\end{enumerate}
then there are no zero modes of the operator $H_f{}{}$ and the functional
$S_f{}(\varphi)$ is strictly positive.
\end{proposition}

In fact, this reduces the problem to the positivity of the matrix $m_f{}$. The
matrix $m_f{}$ has 2 eigenvalues
\be
f^2{}\pm {} |\nabla f|,
\ee
with equal multiplicity, where
\be
|\nabla f|=\sqrt{g^{\mu\nu}\nabla_\mu f
\nabla_\nu f}.
\ee

\begin{proposition}
If the function $f$ satisfies the uniform condition
\be
|\nabla f(x)|<{}{} f^2(x)
\label{416427iga}
\ee
for any $x\in M$, then
the operator $H_f{}{}$ is strictly positive.
\end{proposition}
\proof
Let $W$ be the matrix
\be
W{}=-i[D,f]=\gamma^\mu\nabla_\mu f.
\ee
This matrix is self-adjoint and satisfies the equation
\be
W{}^2=|\nabla f|^2I,
\ee
so, it has two eigenvalues $+|\nabla f|$ and $-|\nabla f|$ with the same multiplicity.
 Let $P_\pm$ be the corresponding projections on the
eigenspaces. Then
\be
\la\varphi, W{}\varphi\ra=|\nabla f|\left(|P_+\varphi|^2-|P_-\varphi|^2\right).
\ee
Since
\be
|P_\pm\varphi|^2\le |\varphi|^2,
\ee
we have
\be
-|\nabla f|\;|\varphi|^2\le \la\varphi, W{}\varphi\ra\le |\nabla f|\;|\varphi|^2.
\ee
Then by using the definition of the matrix $m_f{}$
we have
\be
\left(-{}{}\; |\nabla f|+f^2{}\right)|\varphi|^2\le \la\varphi,m_f{}\varphi\ra
\le \left({}{}\;|\nabla f|+f^2{}\right)|\varphi|^2.
\ee
Therefore,
\be
M_f{}(\varphi)\ge \int_M d\vol\left(-{}{}\;|\nabla f|
+f^2{}\right)|\varphi|^2,
\ee
 and the statement follows, $M_f{}(\varphi)>0$.

By using the decomposition $f=\mu+\tau h$, this condition takes the form
\be
\tau|\nabla h|<(\mu+\tau h)^2;
\ee
if $\mu\ne 0$, it is always satisfied for sufficiently small $\tau$.


\begin{proposition}
Let $f$ be a smooth nonzero function on a compact manifold $M$ without boundary.
Suppose that it satisfies the condition
\be
|\nabla f(x)|<f{}^2(x)
\ee
uniformly on $M$. Then it is either everywhere
positive, $f(x)>0$ for all $x\in M$, or everywhere negative, $f(x)<0$ for all
$x\in M$.
\end{proposition}
\proof
Since the function $f$ is non-zero, then there is a point $x'\in M$ where it is
not-zero. Assume that $f(x')>0$. We consider a geodesic ball $B_{r}(x')$ of
radius $r<r_{\rm inj}(M)$ less than the injectivity radius of the manifold $M$
centered at $x'$. Then we can connect every point $x\in B_r(x')$ in this ball to
the point $x'$ by a geodesic $x(s)$ such that $x(0)=x'$ and $x(t)=x$. We use the
natural parametrization of the geodesic so that $|t|=d(x,x')$ is equal to the
length of the geodesic and the tangent vector has unit norm,
\be
\left|\frac{dx(s)}{ds}\right|^2=1.
\ee

Let $\sigma(x,x')=\frac{1}{2}d^2(x,x')$ be the Ruse-Synge function
equal to one-half the square of the geodesic distance between $x'$
and $x$. Recall that $\sigma(x,x')$ is the solution of the
Hamilton-Jacobi equation
\cite{synge60}
\be
g^{\mu\nu}\nabla_\mu\sigma\nabla_\nu \sigma=2\sigma
\ee
with the boundary conditions
\be
\sigma(x',x')=\nabla_\mu\sigma(x',x')=0,
\ee
and the vector $\sigma_\mu=\nabla_\mu\sigma$ is a tangent vector to the geodesic
at the point $x$. Therefore, the geodesic distance $d=\sqrt{2\sigma}$ satisfies
the equation
\be
g^{\mu\nu}\nabla_\mu d\nabla_\nu d = 1,
\ee
that is, the vector
\be
u_\mu=\nabla_\mu d=\frac{\sigma_\mu}{\sqrt{2\sigma}}
\ee
is the unit tangent vector to the geodesic at the point $x$.

Notice that for any $x\in \supp f$, (where $f(x)\ne 0$),
this condition takes the form
\be
\left|\nabla\left({1\over f}\right)\right|<1.
\ee
Let $\phi$ be a function defined by
\be
\phi(x)=\frac{1}{f(x)}.
\ee
Then
\be
\phi(x')>0
\ee
and for any $x$
\be
|\nabla\phi|^2=
g^{\mu\nu}(x)\nabla_\mu\phi(x)\nabla_\nu\phi(x)<1.
\ee
We evaluate the function $\phi(x(t))$ along the geodesic $x(t)$.
We have
\be
\frac{d\phi(x(s))}{ds}=\frac{dx^\mu(s)}{ds}\nabla_\mu\phi(x(s)).
\ee
Therefore, for any $s\ge 0$
\be
\left|\frac{d\phi(x(s))}{ds}\right|
\le \left|\frac{dx(s)}{ds}\right| |\nabla\phi|
\le 1{} .
\ee
Then
we have
\be
\phi(x(t))=\phi(x(0))+\frac{d\phi(x(s_*))}{ds}t,
\ee
where $0\le s_*\le t$.
Therefore,
\be
|\phi(x)-\phi(x')|\le {} t
\ee
and, by using $t=d(x,x')$, we obtain
\be
\phi(x') - {} d(x,x')\le \phi(x)\le \phi(x') + {} d(x,x').
\ee
Since $f(x')>0$ then
\be
0< \frac{f(x')}{1+f{}(x')d(x,x')}
\le
f(x)
\le
\frac{f(x')}{1-f{}(x')d(x,x')}.
\ee

Finally, by choosing another point $x''$ in the ball $B_r(x')$ we extend this
result to a ball centered at $x''$. Since the manifold $M$ is compact it can be
covered by finitely many geodesic balls where the function is positive.
Therefore, $f$ is positive everywhere.

Similarly, if the function $f$ is negative at some point $f(x')< 0$, then
$-f(x')>0$ and we get
\be
\frac{f(x')}{1+f{}(x')d(x,x')}
\le
f(x)
\le
\frac{f(x')}{1-f{}(x')d(x,x')}
<0 ,
\ee
and the same result follows.


\begin{proposition}
If the function $f$ is nowhere zero, that is, $f(x)>0$ for any $x\in M$ (or
$f(x)<0$ for any $x\in M$) then the Hamiltonian $H_f{}{}$ is strictly positive,
that is, it does not have any zero modes.
\end{proposition}
\proof
Assume that $H_f{}{}\varphi=0$ so that $D_f{}{}\varphi=0$.
Then
\be
(\varphi,f\varphi)=i(\varphi,D\varphi).
\ee
Since the Dirac operator is self-adjoint, the right hand side is imaginary,
and, therefore, $\varphi=0$.

Since $f=\mu+\tau h$, this condition takes the form
\be
\tau |h(x)|<|\mu|,
\ee
and it is satisfied for any smooth
function $h$ and sufficiently small $\tau$
if $\mu\ne 0$.

The converse to this proposition is not true. One can easily construct
a function that is positive everywhere but the condition (\ref{416427iga})
is not satisfied uniformly on $M$.
Suppose that $f(x)>0$  so that
\be
\mu+\tau h(x)>0,
\ee
for any $x\in M$, with $\mu,\tau>0$. Since the average of the function $h$ is equal to zero,
there exists a point $x_0$ such that $h(x_0)=0$.
Suppose it is a nondegenerate point, that is,
$\nabla h(x_0)\ne 0$. Then for sufficiently large constant $\tau$
we have
\be
\tau |\nabla h(x_0)|>\mu^2 .
\ee
Then at the point $x_0$ the condition (\ref{416427iga}) is violated.

This condition is only a {sufficient condition}. It is obvious that it is not
the necessary one, because even though it is a uniform condition but it is only
a local one. Of course, if it is violated in a very small region, then we should
not expect the zero modes show up immediately. If the function $f$ can change
sign then the situation is more complicated.

\section{Properties of the Zero Mode}
\setcounter{equation}{0}

We study the zero modes of the Hamiltonian.
Let $\varphi$ be a non-zero solution of the equation
\be
H_f{}{}\varphi=0.
\ee

\begin{lemma}
Let $J$ be a real vector
\be
J^\mu=\langle{}\varphi,\gamma^\mu\varphi\rangle{}.
\ee
There hold:
\benum
\item
$
||D\varphi||^2 = {} ||f\varphi||^2$,
\item
$(\varphi,f\varphi) = 0$,
\item
$(\varphi,D\varphi) = 0$,
\item
$
\displaystyle
{}{}f|\varphi|^2
=
-{1\over 2}\nabla_\mu J^\mu.
$

\eenum
\end{lemma}
\proof
We have
\be
(\varphi,H_f{}{}\varphi)=||D_f{}{}\varphi||^2=0.
\ee
Therefore, it satisfies the equation
\be
D_f{}{}\varphi=0,
\ee
which means
\be
iD\varphi=f{}\varphi.
\label{612iga}
\ee
In particular, we immediately have
\be
{}||f\varphi||^2 = ||D\varphi||^2,
\ee
and
\be
i(\varphi, D\varphi)=(\varphi,f\varphi).
\ee
This is only possible if both sides vanish,
\be
(\varphi,D\varphi)=0, \qquad (\varphi,f\varphi)=0.
\label{55xx}
\ee

Next, by multiplying eq. (\ref{612iga})
by $\varphi$ {\it pointwise} we get
\be
i\left<\varphi,D\varphi\right>=f{}|\varphi|^2.
\label{614iga}
\ee
Taking the complex conjugate and noting that
the right-hand side here is real we immediately obtain
that $\langle{}\varphi,D\varphi\rangle{}$ is imaginary, that is,
\be
\langle{}\varphi,D\varphi\rangle{}=-\langle{}D\varphi,\varphi\rangle{}.
\ee
Therefore,
\be
{i\over 2}\left(\langle{}\varphi,D\varphi\rangle{}
-\langle{}D\varphi,\varphi\rangle{}\right)=f{}|\varphi|^2.
\ee
Now,
by using the equation
\be
\langle{}\varphi,D\varphi\rangle{}
-\langle{}D\varphi,\varphi\rangle{}
=i\nabla_\mu \langle{}\varphi,\gamma^\mu\varphi\rangle{},
\ee
we obtain
\be
-{1\over 2}\nabla_\mu\langle{}\varphi,\gamma^\mu\varphi\rangle{}
=f{}|\varphi|^2.
\label{57xx}
\ee
Of course, by integrating this equation over $M$ one has
$
(\varphi, f\varphi)=0.
$

\begin{proposition}
There hold:
\benum
\item
The spinor $\varphi$ is not parallel.
\item
The spinor $\varphi$ is not an eigenspinor
of the Dirac operator with a nonzero eigenvalue.
\eenum
\end{proposition}
\proof
If $\nabla\varphi=0$ then $D\varphi=0$, and, by using eq. (\ref{614iga}) we get
$f|\varphi|^2=0$. Since $|\varphi|$ is a non-zero constant this means that $f=0$
everywhere. Next, if $D\varphi=\lambda\varphi$ then by using (\ref{55xx}) we get
$\lambda ||\varphi||^2=0$, and, \ therefore, $\lambda=0$.


Let us denote the nodal set of the function $f$, i.e. the set of
points where $f(x)=0$ by
\be
\Sigma(f)=f^{-1}(0)=\left\{x\in M\;|\; f(x)=0\right\}
\ee
and the corresponding subsets
\bea
M_+(f) &=&f^{-1}(\RR_+)=\left\{x\in M\;|\; f(x)>0\right\},
\\
M_-(f) &=& f^{-1}(\RR_-)=\left\{x\in M\;|\; f(x)<0\right\}.
\eea

We assume that the differential $f_*: T_xM\to \RR$ is surjective at every point
$x\in \Sigma$. Then $\Sigma$ is a $(n-1)$-dimensional (maybe disconnected)
submanifold with a boundary $\partial\Sigma$. We assume that the boundary
$\partial\Sigma$ is smooth and choose the orientation on $\Sigma$ in such a way
that
\be
\partial M_+=\Sigma
=-\partial M_-.
\ee
Recall than the gradient $\nabla f$ is normal to $\Sigma$. We denote by $N$ the
unit normal to the surface $\Sigma$ pointing inside $M_+$ and outside $M_-$.

Then, from (\ref{57xx}) we have
\be
\int_{M_+}d\vol f|\varphi|^2=-\int_{M_-}d\vol f|\varphi|^2.
\ee
Now, let us integrate the eq. (\ref{57xx}) not over
the whole manifold $M$ but over $M_+$ and $M_-$ {\it separately}.
By integrating by parts and using the Stoke's theorem, we get then
\be
{} \int_{M_+}d\vol f\;|\varphi|^2
= -{1\over 2}\int_\Sigma d\vol_\Sigma\; N^\mu J_\mu
= {} \int_{M_-}d\vol |f|\;|\varphi|^2.
\ee
This means that there is a non-zero `flux' of the spinor $\varphi$, more
precisely, the vector $J^\mu$, from the region $M_+$, where $f$ is positive, to
the region $M_-$, where $f$ is negative.

On the other hand, we know that the spectrum of the Hamiltonian $H_f{}{}$ does
not depend on the sign of the function $f$, i.e. it is invariant under the
transformation $f\to -f$. In particular, if $\varphi$ is a zero mode of the
Hamiltonian $H_f{}{}$, then $\Gamma{}\varphi$ is a zero mode of the operator
$H(-f)$. Therefore, if the operator $H_f{}{}$ is strictly positive, then the
operator $H(-f)$ is strictly positive too. Notice that
\be
|\Gamma{}\varphi|
=|\varphi|,\qquad \langle{}\Gamma{}\varphi,\gamma^\mu
\Gamma{}\varphi\rangle{}
=-\langle{}\varphi,\gamma^\mu\varphi\rangle{}.
\ee
Therefore, the `flux' of the spinor $\Gamma{}\varphi$ has the opposite direction,
from $M_-$ to $M_+$, in full correspondence with the fact that for the
spinor $\Gamma{}\varphi$ the regions $M_+$ and $M_-$ interchange their roles
(since we changed the sign of $f$).



There is a basis in which
the Dirac matrices have the off-diagonal block form
\be
\gamma_j=\left(
\begin{matrix}
	0&-i\hat\gamma_j\\
	i\hat\gamma_j&0\\
\end{matrix}\right),
\qquad
\gamma_{n}=\left(
\begin{matrix}
	0&\hat I\\
	\hat I&0\\
\end{matrix}\right),
\label{dirbasis}
\ee
where $\hat\gamma_j$, $j=1,\dots, n-1$, are $2^{m-1}\times 2^{m-1}$ Dirac
matrices in $(n-1)$ dimensions satisfying
\be
\hat\gamma_i\hat\gamma_j+\hat\gamma_j\hat\gamma_i=2 \delta_{ij}\hat I,
\ee
and $\hat\gamma_n=i\hat I$.
Here and everywhere below Latin indices from the middle of the alphabet,
$i,j,k,l,\dots,$ range over $1,\dots, n-1$.
In this basis the chirality operator has the form
\be
\Gamma{}=\left(
\begin{matrix}
	\hat I & 0\\
	0 & -\hat I \\
\end{matrix}\right).
\ee



In the special basis above the Dirac operator has the form
\be
D=\left(\begin{matrix}0&F^*\\{}
	F&0\\
\end{matrix}\right),
\ee
where
\bea
F{} &=& -A+iB,
\\
F^* &=&A+iB,
\eea
where $A$ and $B$ are anti-self-adjoint operators
defined by
\bea
A &=&\hat\gamma^{k}e_k{}^\mu\nabla_{\mu},
\\
B &=& \hat I  e_n{}^\mu\nabla_{\mu}.
\eea
The square of the Dirac operator is
\bea
D^2&=&
\left(\begin{matrix}
F^*F{} &  0\\{}
0 & F{}F^*\\
\end{matrix}\right),
\eea
where
\bea
F^*F{} &=& -A^2-B^2 + i[A,B],
\\
F{}F^* &=& -A^2-B^2 - i[A,B].
\eea


The deformed Dirac operator and the Hamiltonian have the form
\bea
D_f{}{} &=& \left(
\begin{matrix}
i\hat If{} & F^*\\{}
F & i\hat I f{}\\
\end{matrix}\right)
=
\left(\begin{matrix}
i\hat If{} & A+iB \\{}
	-A+iB  & i\hat I f{}\\
\end{matrix}\right),
\label{314iga}
\\[10pt]
H_f{}{} &=&
\left(
\begin{matrix}
F^*F{}+ \hat I f^2 & i C\\{}
-i C^* & F{}F^*+\hat I f^2{}\\
\end{matrix}\right),
\label{315iga}
\eea
where
\bea
C &=& [F^*,f]=[A,f]+i[B,f],
\\
C^* &=& -[F{},f]=[A,f]-i[B,f].
\eea


By using the spiral decomposition,
$\varphi=\varphi_+\oplus\varphi_-$, and the form (\ref{314iga})
of the deformed Dirac operator, the equation $D_f{}{}\varphi=0$
gives
\bea
F{} \varphi_+ +if{}\varphi_- &=&0,
\label{622iga}\\
F^*\varphi_- +if{} \varphi_+ &=&0.
\label{623iga}
\eea
Note, also that if $f(x)\ne 0$ then we have
\bea
\left(F^*\frac{1}{f}F{}+{} f\right)\varphi_+ &=&0,
\\
\left(F{}\frac{1}{f}F^*+{} f\right)\varphi_- &=&0.
\eea
therefore,
\bea
\left(\int_{M_+}+\int_{M_-}\right)
d\vol\left\{\frac{1}{f}\left|F{}\varphi_+\right|^2+{}
f|\varphi_+|^2\right\}
&=&0,
\\
\left(\int_{M_+}+\int_{M_-}\right) d\vol\left\{\frac{1}{f}
\left|F^*\varphi_-\right|^2+{} f|\varphi_-|^2\right\}&=&0.
\eea
therefore, if the function $f$ is positive (or negative)
then $\varphi_+=\varphi_-=0$.

Similarly, by using the form (\ref{315iga})
of the Hamiltonian,
we obtain
\bea
H_+ \varphi_++i{} C\varphi_- &=&0,
\\
-i{} C^*\varphi_+ + H_-\varphi_- &=&0,
\eea
where
\bea
H_+ &=& F^*F{}+f^2{},
\\
H_-&=& F{}F^*+f^2{}.
\eea
Note that for a non-zero function $f$, the operators
$H_+$ and $H_-$
are positive,
Therefore, we have
\bea
\varphi_- &=& i{}H_-^{-1} C^*\varphi_+,
\\
\varphi_+ &=& -i{}H_+^{-1} C\varphi_-,
\eea
which gives the equations
\bea
\left\{H_+ -{} CH_-^{-1}C^*\right\}\varphi_+ &=& 0,
\\
\left\{H_- -{} C^*H_+^{-1}C\right\}\varphi_- &=& 0.
\eea
This means, in particular,
\bea
\Big|\Big|\sqrt{H_+}\varphi_+\Big|\Big|^2  &=& \Big|\Big|\frac{1}{\sqrt{H_-}}C^*\varphi_+\Big|\Big|^2,
\\
\Big|\Big|\sqrt{H_-}\varphi_-\Big|\Big|^2  &=& \Big|\Big|\frac{1}{\sqrt{H_+}}C\varphi_-\Big|\Big|^2.
\eea

\begin{proposition}
Suppose that the operator $A$ has a zero mode $\psi_{f}$ that satisfies the
equations
\be
A\psi_{f} = (B+f{})\psi_{f} =0,
\ee
Then the spinors
\bea
\varphi_1
&=& \left(\begin{matrix}
\psi_{f}\\{}
\psi_{f}\\
\end{matrix}\right),
\\
\varphi_2
&=&
\left(\begin{matrix}
\psi_{-{f}}\\{}
-\psi_{-{f}}\\
\end{matrix}\right),
\eea
are the zero modes of the Hamiltonian $H_f{}{}$,
\be
H_f{}{}\varphi_1=H_f{}{}\varphi_2=0.
\ee
\end{proposition}
\proof
First of all, since the operators $A$ and $B$ do not depend on
${f}$, we notice that the spinor
$\psi_{-{f}}$ satisfies the equations
\be
A\psi_{-{f}} = (B-f{})\psi_{-{f}} =0.
\ee
By using the form of the operator $F^*=A+iB$,
eqs. (\ref{622iga}) and (\ref{623iga}) take the form
\bea
(B+iA) \varphi_+ +f{}\varphi_- &=&0,
\\
(B-iA)\varphi_- +f{} \varphi_+ &=&0,
\eea
By adding and subtracting these equations we get
\bea
(B+f{}) \phi + iA\chi &=&0,
\\
iA\phi +(B-f{})\chi &=&0,
\eea
where
\be
\phi=\frac{1}{2}\left(
\varphi_++\varphi_-\right),
\qquad
\chi=\frac{1}{2}\left(
\varphi_+-\varphi_-\right).
\ee
By combining these equations we also get a useful equation
\bea
\left\{(B-f{})(B+f{}) + A^2\right\}\phi &=& i[A,B-f{}]\chi,
\\
\left\{(B+f{})(B-f{}) + A^2\right\}\chi &=& i[A,B+f{}]\phi.
\eea

These equations are satisfied if
\be
\phi=\psi_{f}
\qquad\mbox{and}\qquad
\chi=0,
\ee
that is, $\varphi_-=\varphi_+=\psi_{f}$,
or if
\be
\psi=0
\qquad\mbox{and}\qquad
\chi=\psi_{-{f}},
\ee
that is, $\varphi_+=-\varphi_-=\psi_{-{f}}$.

We will give an example of such a solution in the next section.

\section{Example of a Zero Mode}
\setcounter{equation}{0}

We provide a counterexample demonstrating that for an arbitrary function $f$ the
Hamiltonian is not necessarily positive, that is, it could have zero modes.

Let $N$ be a closed $(n-1)$-dimensional manifold (with $n=2m$ being even) with
local coordinates $\hat x^i$, $i=1,\dots, n-1$, with a Riemannian metric
\be
dl^2=\hat g_{ij}(\hat x)d\hat x^i\;d\hat x^j.
\ee
We adopt a convention that then Latin indices from the middle of the alphabet
range over $1, \dots, n-1$. Let $r$, $0\le r\le 2\pi$, be a coordinate of a unit
circle $S^1$ and $M=N\times S^1$ be a product manifold with the metric
\be
ds^2=dl^2+dr^2.
\ee
This defines the orthonormal frame
\be
\sigma^{(i)}=\sigma^{(i)}{}_j(\hat x)d\hat x^j,
\qquad
\sigma^{(n)}=dr,
\ee
\be
e_{(i)}=e_{(i)}{}^j(\hat x)\hat \partial_j
\qquad e_{(n)}=\partial_r.
\ee
The only non-zero components of the spin connection are
$\omega_{(i)(j)(k)}(\hat x)$. We use Latin letters in parenthesis
to distinguish the frame indices from the coordinate indices.

Therefore, the Dirac operator
takes the form
\bea
D &=& \tilde D +i\gamma_n\partial_r,
\eea
where
\be
\gamma_n=\left(\begin{matrix} 0 & \hat I\\{}
	\hat I & 0\\
\end{matrix}\right)
\ee
and
\be
\tilde D=i\gamma^{(j)}e_{(j)}{}^k\hat\nabla_k.
\ee
In the special basis
(\ref{dirbasis})
the operator $\tilde D$ takes the form
\be
\tilde D=\left(\begin{matrix} 0 & -i\hat D\\{}
	i\hat D&0\\
\end{matrix}\right),
\ee
where
\bea
\hat D &=& i\hat\gamma^j e_{(j)}{}^k\hat\nabla_k
\eea
is nothing but the Dirac operator on the manifold $N$.

We assume that the operator $\hat D$ has a nontrivial kernel,
that is, it has zero modes. For example, the manifold $N$
could be the product $N=\Sigma\times S^1$, where
$\Sigma$ is an even-dimensional manifold with a non-zero
index of the Dirac operator.

Therefore, the Dirac operator on the product manifold $M=N\times S^1$ is
\be
D=\left(\begin{matrix} 0 &  -i\hat D +i\hat I\partial_r\\{}
i\hat D +i\hat I\partial_r &0\\
\end{matrix}\right).
\ee


Let $f=f(r)$ be a smooth function on $S^1$
(which is, of course, periodic and is constant on $N$)
normalized by
\be
||f||^2_M=\vol(N) \int_0^{2\pi}dr\; |f|^2=1,
\ee
We decompose the function $f$ by separating the constant term
\be
f=\mu+\tau h,
\label{712iga}
\ee
where
\be
\mu=\frac{1}{\vol(N)}\frac{1}{2\pi}\int_0^{2\pi}dr f(r),
\ee
and $h$ is a periodic function such that
\be
\int_{0}^{2\pi}dr h(r)=0.
\ee
Then the function $h$ is normalized by
\be
\int_0^{2\pi}dr\; |h(r)|^2=\frac{1}{\vol(N)}.
\ee
Further, we define the function
$\omega=\omega(r)$ by
\be
\omega(r) = \int_0^r dt\;h(t),
\ee
so that $h(r)=\omega'(r)$;
obviously, $\omega(r)$ is also periodic.

Then the deformed Dirac operator is
\bea
D_f{}{} &=& \tilde D + i \gamma_n \partial_r
+iIf{}(r)
\nonumber\\
&=&\left(
\begin{matrix}
i\hat I f{} &  -i\hat D + i\hat I\partial_r\\{}
i\hat D +i\hat I\partial_r & i\hat I f{}\\
\end{matrix}\right).
\eea
and the Hamiltonian operator is
\bea
H_f{}{} &=&
I\left[\hat D^2-\partial_r^2 +f^2{}(r)\right]
-\gamma_nf{}'(r).
\eea
Then by using the spiral decomposition
 the functional $S_f{}(\varphi)$
has the form
\bea
S_f{}(\varphi) &=&
\int\limits_N d\vol_N
\int\limits_0^{2\pi}dr
\Biggl\{
\Big|\hat D\varphi_++\partial_r\varphi_+ +f{}\varphi_-\Big|^2
+\Big|\hat D\varphi_--\partial_r\varphi_- -f{}\varphi_+\Big|^2
\Biggr\}
\nn\\
&=&\int\limits_N d\vol_N
\int\limits_0^{2\pi}dr
\Biggl\{
|\hat D\varphi_+|^2+|\hat D\varphi_-|^2
+|\partial_r\varphi_+|^2+|\partial_r\varphi_-|^2
\nonumber\\
&&
+f{}(r)\Bigl[\la \partial_r\varphi_+,\varphi_-\ra
+\la \varphi_-,\partial_r\varphi_+\ra
+\la \partial_r\varphi_-,\varphi_+\ra
+\la \varphi_+,\partial_r\varphi_-\ra
\Bigr]
\nn\\
&&+f^2{}(r)\left(|\varphi_+|^2 +|\varphi_-|^2\right)
\Biggr\};
\eea

\begin{proposition}
Suppose that the function $f=f(r)$ has the zero average over the circle $S^1$
\be
\int_{0}^{2\pi}dt f(t)=0.
\ee
Suppose that the Dirac operator $\hat D$ has a zero mode
$\psi_0$
 on the manifold $N$,
\be
\hat D\psi_0=0.
\ee
Then the spinors
\bea
\varphi_1(\hat x,r)
&=& \exp\left[-\tau \omega(r)\right]
\left(\begin{matrix}
\psi_0(\hat x)\\{}
\psi_0(\hat x)\\
\end{matrix}\right),
\\
\varphi_2(\hat x,r)
&=&
\exp\left[\tau \omega(r)\right]
\left(\begin{matrix}
\psi_0(\hat x)\\{}
-\psi_0(\hat x)\\
\end{matrix}\right),
\eea
are zero modes of the Hamiltonian $H_f{}{}$ on
the product manifold $M=N\times S^1$,
\be
H_f{}{}\varphi_1=H_f{}{}\varphi_2=0,
\ee
with the norms
\be
||\varphi_{1,2}||^2=
A_{1,2}||\psi_0||_N^2,
\ee
where
\be
A_{1,2} = 2\int_0^{2\pi}dr\;\exp\left[\mp 2\tau\omega(r)\right].
\ee
\end{proposition}
\proof
The equation for the zero mode of the deformed Dirac operator $D_f{}{}$ is
\be
\left\{\cD+I\partial_r+\gamma_n f{}
\right\}\varphi=0,
\label{714igax}
\ee
where
\be
\cD=-i\gamma_n\tilde D
=\left(\begin{matrix}
\hat D & 0\\{}
0 & -\hat D\\
\end{matrix}\right)
\ee
or
\bea
\hat D\varphi_+ +\partial_r \varphi_+ + f{}\varphi_- &=&0,
\\
-\hat D\varphi_- +\partial_r \varphi_- + f{}\varphi_+ &=&0.
\eea

By adding and subtracting these equations we get
\bea
\hat D\chi+(\partial_r+f{}) \psi  &=&0,
\\
\hat D\psi+(\partial_r- f{}) \chi  &=&0,
\eea
where
\be
\psi=\frac{1}{2}(\varphi_++\varphi_-),
\qquad
\chi=\frac{1}{2}(\varphi_+-\varphi_-).
\ee
Since the operator $\hat D$ commutes with the operator $\partial_r$ and the
function $f$, this gives two separate second-order equations
\bea
\left\{\hat D^2-(\partial_r-f{})(\partial_r+ f{}) \right\}\psi &=&0,
\\
\left\{\hat D^2-(\partial_r+f{})(\partial_r- f{}) \right\}\chi &=&0.
\eea
by multiplying these equations by $\psi$ and $\chi$ correspondingly we obtain
\bea
\int\limits_N d\vol_N
\int\limits_0^{2\pi}dr
\Biggl\{
|\hat D\psi|^2 + |(\partial_r+f{})\psi|^2
\Biggr\} &=&0,
\\
\int\limits_N d\vol_N
\int\limits_0^{2\pi}dr
\Biggl\{
|\hat D\chi|^2 + |(\partial_r-f{})\chi|^2
\Biggr\} &=&0.
\eea
Therefore, they have to
be the zero modes of the Dirac operator on the manifold $N$,
\be
\hat D\psi = \hat D\chi=0,
\ee
and satisfy the first-order equations
\bea
(\partial_r+f{})\psi &=& 0,
\\
(\partial_r-f{})\chi &=& 0.
\eea
By using the decomposition (\ref{712iga}) we get the solutions
\bea
\psi(\hat x,r) &=& \exp\left[-\mu r-\tau \omega(r)\right]\psi_0(\hat x),
\\
\chi(\hat x,r) &=& \exp\left[\mu r+\tau \omega(r) \right]\chi_0(\hat x),
\eea
where $\psi_0(\hat x)$ and $\chi_0(\hat x)$ are some zero modes of the Dirac
operator $\hat D$. Note that for $\mu\ne 0$ these solutions are not periodic and
are not genuine zero modes. However, for $\mu=0$ they give the zero mode of the
deformed Dirac operator $D_f{}{}$ (and, therefore, of the Hamiltonian $H_f{}{}$)
for an arbitrary function $h(r)=\omega'(r)$,
\bea
\varphi_+(\hat x,r) &=&
\exp\left[-\tau\omega(r)\right]\psi_0(\hat x)
+
\exp\left[\tau \omega(r)\right]\chi_0(\hat x),
\\
\varphi_-(\hat x,r) &=&
\exp\left[-\tau \omega(r)\right]\psi_0(\hat x)
-
\exp\left[\tau \omega(r)\right]\chi_0(\hat x),
\eea
that is,
\be
\varphi=\varphi_1+\varphi_2,
\ee
where
\be
\varphi_1(\hat x,r)
=\exp\left[-\tau \omega(r)\right]
\left(\begin{matrix}
\psi_0(\hat x)\\{}
\psi_0(\hat x)\\
\end{matrix}\right),
\qquad
\varphi_2(\hat x,r)
=
\exp\left[\tau \omega(r)\right]
\left(\begin{matrix}
\chi_0(\hat x)\\{}
-\chi_0(\hat x)\\
\end{matrix}\right).
\ee
The norms of these solution are
\be
||\varphi_{1}||^2=
A_{1}||\psi_0||_N^2,
\qquad
||\varphi_{2}||^2=
A_{2}||\chi_0||_N^2,
\ee
where
\bea
A_{1,2}=2\int_0^{2\pi}dr\;\exp\left[\mp 2\tau \omega(r)\right].
\eea

\section{Conclusion}
\setcounter{equation}{0}

The primary goal of this paper was to study the kernel of a deformed Dirac operator
related to the zero energy states of a corresponding Hamiltonian acting on spinor fields
over a closed Riemannian manifold.
First, we obtained some sufficient conditions on the deformation function that ensure
the absence of the zero modes and the positivity of the Hamiltonian.
Then we showed that these conditions are not necessary by constructing
an explicit counterexample of a deformation function on a product manifold
that leads to a non-trivial kernel of the deformed Dirac operator.



\begin{thebibliography}{99}

\bibitem{avramidi00}
 I. G. Avramidi, {\it Heat Kernel and Quantum Gravity}, Berlin: Springer, 2000

\bibitem{avramidi05}
I. G. Avramidi,
{\it Dirac operator in matrix geometry}, Int. J. Geom. Meth. Mod. Phys.
{\bf 2} (2005) 227-264


\bibitem{berline92}
N. Berline, E. Getzler and M. Vergne,
{\it Heat Kernels and Dirac Operators}, Berlin, Springer-Verlag, 1992

\bibitem{cycon08}
H. L. L. Cycon, R. G. Froese, W. Kirsch and B. Simon,
{\it Schr\"odinger Operators: With Applications to Quantum Mechanics and Global Geometry},
(Theoretical and Mathematical Physics), Springer; 2008

\bibitem{fefferman81}
C. Fefferman and D. Phong, {\it The uncertainty principle and sharp G\aa rding inequalities},
Commun. Pure Appl. Math. {\bf 34} (1981) 285-331

\bibitem{friedrich00}
T. Friedrich, {\it Dirac Operators in Riemannian Geometry}, Graduate
Studies in Mathematics, Vol. 25, Providence, Rhode Island, American
Mathematical Society, 2000

\bibitem{froehlich00}
J. Fr\"ohlich, G.M. Graf, D. Hasler, J. Hoppe and S.-T. Yau,
{\it Asymptotic form of zero energy wave functions in supersymmetric
matrix models},
Nucl.Phys. {\bf B567} (2000) 231-248

\bibitem{froehlich98}
J. Fr\"ohlich and J. Hoppe,
{\it On zero-mass ground states in super-membrane matrix models},
Comm. Math. Phys. {\bf 191} (1998) 613-626

\bibitem{dewit89}
 B. de Wit, M. L\"uscher and H. Nicolai, {\it The supermembrane is unstable}, Nucl. Phys. B
 320 (1989) 135-159

\bibitem{gilkey95}
 P. B. Gilkey, {\it Invariance Theory, the Heat Equation and the Atiyah-Singer
Index Theorem}, CRC Press, Boca Raton, 1995

\bibitem{graf01}
G. M. Graf, D. Hasler and  J. Hoppe,
{\it No zero energy states for the supersymmetric
$x^2y^2$ potential}, arXiv:math-ph/0109032

\bibitem{simon83}
B. Simon, {\it Some quantum operators with discrete spectrum but classically continuous
spectrum}, Ann. Phys. {\bf 146} (1983) 209-220

\bibitem{synge60}
J. L. Synge, {\it Relativity: The General Theory}, Amsterdam: North-Holland,
1960

\bibitem{tong} D. Tong,
{\it Supersymmetric Quantum Mechanics},
Cambridge University,
https://www.damtp.cam.ac.uk/user/tong/susy/susy.pdf

\bibitem{zhelnorovich19}
 V. A. Zhelnorovich, {\it Theory of Spinors and Its Application in Physics
and Mechanics},
 Springer, 2019


\end{thebibliography}
\end{document}